\documentclass[12pt]{article}
\usepackage{a4,amssymb,epsf}

\newcommand{\binom}[2]{\left(\begin{array}{@{}c@{}}#1\\#2\end{array}\right)}

\newcommand{\ZZ}{\mathbb Z}
\newcommand{\EE}{\mathbb E}

\begin{document}

\setlength{\baselineskip}{1.1\baselineskip}
\centerline{\Large\bf Coordination sequences 
             for root lattices and related graphs}\vspace*{8mm}

\centerline{\large Michael Baake\footnotemark[1]\ and 
            Uwe Grimm\footnotemark[2]}

\begin{center}
\footnotemark[1]\
\begin{minipage}[t]{100mm}{
                Institut f\"{u}r Theoretische Physik,
                Universit\"{a}t T\"{u}bingen,\\
                Auf der Morgenstelle 14,
                D--72076 T\"{u}bingen, Germany}
\end{minipage}\\[2mm]
\footnotemark[2]\
\begin{minipage}[t]{100mm}{
                Institut f\"{u}r Physik,
                Technische Universit\"{a}t Chemnitz,\\
                D--09107 Chemnitz, Germany}
\end{minipage}
\end{center}\vspace*{5mm}

\centerline{Dedicated to Ted Janssen on the occasion of his 60th birthday}

\begin{quote}
The coordination sequence $s_{\Lambda}^{}(k)$ of a graph $\Lambda$ 
counts the number of its vertices which have distance $k$ from 
a given vertex, where the distance between two vertices is defined 
as the minimal number of bonds in any path connecting them. 
For a large class of graphs, including in particular the classical 
root lattices, we present the coordination sequences and their
generating functions, summarizing and extending recent results
of Conway and Sloane \cite{ConSlo}.
\end{quote}

\section*{Introduction}

Discrete versions of physical models are usually based on graphs, particularly
on periodic lattices. For instance, a lattice may serve as an abstraction 
of the regular arrangement of atoms in a crystalline solid, and the
physical model introduces suitable degrees of freedom associated
to the vertices or the edges of the graph, depending on the type of
physical property one intends to study. Conversely, for a given
lattice model describing some interesting physical situation,
one might be interested to understand the influence of the underlying
graph on the physical properties of the system.

For example, an important class of lattice models are classical
and quantum spin models intended to describe magnetic ordering,
the famous Ising model being the simplest and most thoroughly
studied member of this group. For these models, one is mainly
interested in their critical properties, i.e., the behaviour of 
physical quantities at and in the vicinity of the phase transition
point where magnetic ordering occurs. In many cases, these critical
properties are ``universal'' in the sense that they do not depend
on the details of the particular model under consideration, but only
on a number of rather general features such as the 
space dimension, the symmetries and the range of the interactions. 

In contrast, the precise location of the critical point 
(the critical temperature) depends sensitively on the underlying graph.
It has been demonstrated recently \cite{GalMau1,GalMau2} that the location 
of the critical points for Ising and percolation models on several 
lattices can be well approximated by empirical functions involving the
dimension and the coordination number of the lattice. On the other hand,
these quantities alone cannot completely determine the critical point 
as can be seen from data obtained for different graphs with identical
dimension and (mean) coordination number \cite{Bri,Laby,LY}. 
This poses the question
how to include more detail of the lattice in order to improve the
approximation. In our view, it is the natural approach to investigate
higher coordination numbers (i.e., the number of next-nearest
neighbours and so on) of the lattice and their influence
on the physical properties of the model.

With this in mind, we started to analyze the coordination sequences of
various graphs, and, in particular, the classical root lattices 
\cite{CS,Humphreys}. Apart from some
numerical investigation \cite{Keefe91,Keefe95}, this did not seem to
have attracted a lot of research. However, when we finished our calculations
and started to work on the proofs, we became aware of recent results of
Conway and Sloane \cite{ConSlo} where the problem is solved 
for the root lattices $A_n$ ($n\geq 1$), $D_n$ ($n\geq 4$),
$E_6$, $E_7$, and $E_8$, together with proofs for most of the 
results. (The corresponding sequences are not contained 
in \cite{SloPlo95}, but have been added to \cite{Slo}.)
Not treated, however, are the periodic graphs obtained from the root
systems $B_n$ ($n\geq 2$), $C_n$ ($n\geq 2$), $F_4$, and $G_2$. They
do not result in new {\em lattices\/} (seen as the set of points
reached by integer linear combinations of the root vectors), but they
do result in different {\em graphs\/}, because they have rather different
connectivity patterns. We thus call them {\em root graphs\/} from now on.

In what follows, we present the results on the coordination sequences and
their generating functions in a concise way, including some of the material
of \cite{ConSlo}, but omitting proofs. The latter, in many cases, follow
directly from \cite{ConSlo} or can be traced back to it --- with two 
exceptions mentioned explicitly later on.

\section*{Preliminaries and general setup}

The calculation of the coordination sequence of a lattice first means to
specify the corresponding graph, i.e., to specify who is neighbour of whom
in the lattice. In the simplest example of all, the lattice $\ZZ$, 
each lattice point has precisely two neighbours, 
one to the left and one to the right.
Consequently, the number $s_{\ZZ}^{}(k)$ of $k$th neighbours is 
$s^{}_{\ZZ}(0)=1$ and $s^{}_{\ZZ}(k)=2$ for $k\ge 1$, 
with generating function
\begin{equation}
     S^{}_{\ZZ}(x) \; = \; 
     \sum_{k=0}^{\infty}\: s^{}_{\ZZ}(k)\; x^k \; = \; \frac{1+x}{1-x} \, ,
\end{equation}
compare \cite{Wilf} for elementary background material on this type 
of approach.
If we combine two lattices $\Lambda_1,\Lambda_2$ 
in Euclidean spaces $\EE_1,\EE_2$,
respectively, to the direct sum 
$\Lambda_1\oplus\Lambda_2$ in $\EE_1\oplus\EE_2$,
together with the rule that $x=(x_1,x_2)$ is neighbour of $y=(y_1,y_2)$ in
$\Lambda_1\oplus\Lambda_2$ if and only if $x_1$ is neighbour of $y_1$ in
$\Lambda_1$ {\em and} $x_2$ is neighbour of $y_2$ in $\Lambda_2$, 
the new generating function is a product:
\begin{eqnarray}
S^{}_{\Lambda_1\oplus\Lambda_2}(x) 
     & = & \sum_{k=0}^{\infty}\: s^{}_{\Lambda_1\oplus\Lambda_2}(k)\; x^k 
           \nonumber \\
     & = & \sum_{m=0}^{\infty} \sum_{\ell=0}^{m}\: 
           s^{}_{\Lambda_1}(\ell)\: s^{}_{\Lambda_2}(m-\ell)\; x^m 
           \nonumber \\
     & = & S^{}_{\Lambda_1}(x) \cdot S^{}_{\Lambda_2}(x) \, . 
\end{eqnarray} 

A direct application to the situation of the cubic lattice $\ZZ^n_{}$
immediately gives its generating function
\begin{equation}
    S^{}_{\ZZ^n}(x) \; = \; \left( \frac{1+x}{1-x} \right)^n 
  \; = \; \frac{1}{{(1-x)}^n}\:\sum_{k=0}^{n}\binom{n}{k} x^k
\end{equation}
which (accidentally) coincides with its $\theta$-function \cite{CS}.
Similarly, if we know the generating functions for certain lattices, we can
extend them to all direct sums of this type. It is thus reasonable to take
a closer look at the root lattices (see \cite{CS} for definition and
background material and \cite{Humphreys} for details on the underlying root
systems and their classification). In view of the previous remark,
it is sufficient to restrict to the simple root lattices 
which are characterized by connected Dynkin diagrams \cite{Humphreys,CS}.
The corresponding graphs are obtained by the rule that a lattice point $x$
has all other lattice points as neighbours that can be reached by a 
root vector. Note that, due to this rule, {\em all\/} root systems will
appear. As an example, consider $A_2$ and $G_2$: they define the same
root lattice, but different graphs and hence different coordination
sequences, see Figure~1. Also, $F_4$ defines the same lattice as $D_4^*$,
the dual of $D_4$ and equivalent to it as a lattice, but not the same
graph. Similarly, $B_n$ (for which the root lattice is just $\ZZ_{}^n$)
and $C_n$ (whose root lattice coincides with that of $D_n$) define 
different graphs for $n\geq 3$, while those of
$B_2$ and $C_2$ are equivalent (they yield a square lattice with points
connected along the edges and the diagonals of the squares).

\begin{figure}[ht]
\centerline{\epsfxsize=0.7\textwidth \epsfbox{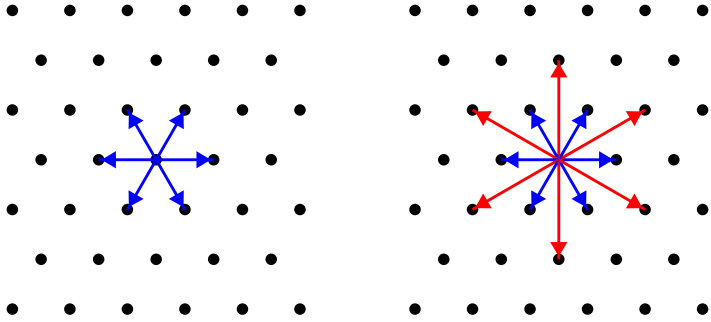}}
\caption{\small The vector stars of the root systems $A_2$ (left) 
and $G_2$ (right).
They define neighbouring points in the corresponding root graphs.
\label{fig1}}
\end{figure}

In all examples to be discussed below, the generating function is
of the form
\begin{equation} \label{poly}
    S_{\Lambda}^{}(x) \; = \; \frac{P_{\Lambda}^{}(x)}{(1-x)^n}
\end{equation}
where $\Lambda$ is a lattice in $n$-dimensional Euclidean space and
$P_{\Lambda}^{}(x)$ is an integral polynomial of degree $n$
(for a proof of this statement for the root lattices, see \cite{ConSlo};
the remaining cases rest upon the proper generalization of the concept
of well-roundedness to root graphs).
It is therefore sufficient to list the polynomials in the numerator
of (\ref{poly}) for the lattices and graphs under consideration.

\section*{Results}

Although we shall give the generating functions below,
the explicit values of the coordination numbers $s^{}_{\Lambda}(k)$ 
of root graphs $\Lambda$ in dimension $n\le 8$ are, for convenience,
shown in Table~\ref{tab1l} for $1\le k\le 10$. Note that, by definition,
we set $s^{}_{\Lambda}(0)=1$ in all cases. 

\begin{table}[ht]
\caption{\small
First coordination numbers of root graphs of dimension $n\le 8$}
\begin{small}
\begin{center}
\label{tab1l}
\makebox[0pt]{
\begin{tabular}{lrrrrrrrrrr}
\hline\hline
\multicolumn{1}{c}{$\rule[-1.3ex]{0ex}{4ex}\Lambda$} &
\multicolumn{1}{c}{$s^{}_{\Lambda}(1)$} &
\multicolumn{1}{c}{$s^{}_{\Lambda}(2)$} &
\multicolumn{1}{c}{$s^{}_{\Lambda}(3)$} &
\multicolumn{1}{c}{$s^{}_{\Lambda}(4)$} &
\multicolumn{1}{c}{$s^{}_{\Lambda}(5)$} &
\multicolumn{1}{c}{$s^{}_{\Lambda}(6)$} &
\multicolumn{1}{c}{$s^{}_{\Lambda}(7)$} &
\multicolumn{1}{c}{$s^{}_{\Lambda}(8)$} &
\multicolumn{1}{c}{$s^{}_{\Lambda}(9)$} &
\multicolumn{1}{c}{$s^{}_{\Lambda}(10)$}
\\
\hline
$A_1$ \rule[0ex]{0ex}{2.7ex}
      & 2 & 2 & 2 & 2 & 2 & 2 & 2 & 2 & 2 & 2 \\
$A_2$ & 6 & 12 & 18 & 24 & 30 & 36 & 42 & 48 & 54 & 60 \\
$A_3$ & 12 & 42 & 92 & 162 & 252 & 362 & 492 & 642 & 812 & 1002 \\
$A_4$ & 20 & 110 & 340 & 780 & 1500 & 2570 & 4060 & 6040 & 8580 & 11750 \\
$A_5$ & 30 & 240 & 1010 & 2970 & 7002 & 14240 & 26070 & 44130 & 70310 
      & 106752 \\
$A_6$ & 42 & 462 & 2562 & 9492 & 27174 & 65226 & 137886 & 264936 & 472626
      & 794598 \\
$A_7$ & 56 & 812 & 5768 & 26474 & 91112 & 256508 & 623576 & 1356194 & 2703512 
      & 5025692 \\ 
$A_8$ \rule[-1.3ex]{0ex}{1.3ex}
      & 72 & 1332 & 11832 & 66222 & 271224 & 889716 & 2476296 & 6077196 
      & 13507416 & 27717948 \\
\hline
$B_2$\rule[0ex]{0ex}{2.7ex}
      & 8 & 16 & 24 & 32 & 40 & 48 & 56 & 64 & 72 & 80 \\
$B_3$ & 18 & 74 & 170 & 306 & 482 & 698 & 954 & 1250 & 1586 & 1962 \\
$B_4$ & 32 & 224 & 768 & 1856 & 3680 & 6432 & 10304 & 15488 & 22176 & 30560 \\
$B_5$ & 50 & 530 & 2562 & 8130 & 20082 & 42130 & 78850 & 135682 & 218930
      & 335762 \\
$B_6$ & 72 & 1072 & 6968 & 28320 & 85992 & 214864 & 467544 & 918080 & 1665672
      & 2838384 \\
$B_7$ & 98 & 1946 & 16394 & 83442 & 307314 & 907018 & 2282394 & 5095650 
      & 10368386 & 19594106 \\
$B_8$ \rule[-1.3ex]{0ex}{1.3ex}
      & 128 & 3264 & 34624 & 216448 & 954880 & 3301952 & 9556160 
      & 24165120 & 54993792 & 115021760 \\
\hline
$C_2$ \rule[0ex]{0ex}{2.7ex}
      & 8 & 16 & 24 & 32 & 40 & 48 & 56 & 64 & 72 & 80 \\
$C_3$ & 18 & 66 & 146 & 258 & 402 & 578 & 786 & 1026 & 1298 & 1602 \\
$C_4$ & 32 & 192 & 608 & 1408 & 2720 & 4672 & 7392 & 11008 & 15648 & 21440 \\
$C_5$ & 50 & 450 & 1970 & 5890 & 14002 & 28610 & 52530 & 89090 & 142130 
      & 216002 \\
$C_6$ & 72 & 912 & 5336 & 20256 & 58728 & 142000 & 301560 & 581184 & 1038984
      & 1749456 \\
$C_7$ & 98 & 1666 & 12642 & 59906 & 209762 & 596610 & 1459810 & 3188738 
      & 6376034 & 11879042 \\
$C_8$ \rule[-1.3ex]{0ex}{1.3ex}
      & 128 & 2816 & 27008 & 157184 & 658048 & 2187520 & 6140800 
      & 15158272 & 33830016 & 69629696 \\
\hline
$D_4$ \rule[0ex]{0ex}{2.7ex}
      & 24 & 144 & 456 & 1056 & 2040 & 3504 & 5544 & 8256 & 11736 & 16080 \\
$D_5$ & 40 & 370 & 1640 & 4930 & 11752 & 24050 & 44200 & 75010 & 119720
      & 182002 \\
$D_6$ & 60 & 792 & 4724 & 18096 & 52716 & 127816 & 271908 & 524640 & 938652
      & 1581432 \\
$D_7$ & 84 & 1498 & 11620 & 55650 & 195972 & 559258 & 1371316 & 2999682 
      & 6003956 & 11193882 \\
$D_8$ \rule[-1.3ex]{0ex}{1.3ex}
      & 112 & 2592 & 25424 & 149568 & 629808 & 2100832 & 5910288 
      & 14610560 & 32641008 & 67232416 \\
\hline
$E_6$ \rule[0ex]{0ex}{2.7ex}
      & 72 & 1062 & 6696 & 26316 & 77688 & 189810 & 405720 & 785304 
      & 1408104 & 2376126 \\
$E_7$ & 126 & 2898 & 25886 & 133506 & 490014 & 1433810 & 3573054 
      & 7902594 & 15942206 & 29896146 \\
$E_8$ \rule[-1.3ex]{0ex}{1.3ex}
      & 240 & 9120 & 121680 & 864960 & 4113840 & 14905440 & 44480400 
      & 114879360 & 265422960 & 561403680 \\
\hline
$F_4$ \rule[-1.3ex]{0ex}{4ex}
      & 48 & 384 & 1392 & 3456 & 6960 & 12288 & 19824 & 29952 & 43056 
      & 59520 \\
\hline
$G_2$ \rule[-1.3ex]{0ex}{4ex}
      & 12 & 30 & 48 & 66 & 84 & 102 & 120 & 138 & 156 & 174 \\
\hline\hline
\end{tabular}}
\end{center}
\end{small}
\end{table}

The coordinator polynomials $P_{\Lambda}^{}(x)$ of the root graphs 
belonging to the four infinite series turn out to be given by
\begin{eqnarray}
P_{A_n}^{}(x) & = & 
\sum_{k=0}^{n}\; {\binom{n}{k}}^{2}\; x^k \\
P_{B_n}^{}(x)
& = & \frac{1}{2}\left[{(1+\sqrt{x})}^{2n+1}+{(1-\sqrt{x})}^{2n+1}\right]
\; -\; 2nx{(1+x)}^{n-1}\nonumber\\*
& = & \sum_{k=0}^{n}\;\left[\,
\binom{2n+1}{2k}\: -\: 2k\:\binom{n}{k}\,\right]\; x^k \\
P_{C_n}^{}(x) 
& = & \frac{1}{2}\left[{(1+\sqrt{x})}^{2n}+{(1-\sqrt{x})}^{2n}\right] 
\;\; = \;\; \sum_{k=0}^{n}\;\binom{2n}{2k}\; x^k \\
P_{D_n}^{}(x) 
& = & \frac{1}{2}\left[{(1+\sqrt{x})}^{2n}+{(1-\sqrt{x})}^{2n}\right]
\; -\; 2nx{(1+x)}^{n-2} \nonumber\\*
& = & \sum_{k=0}^{n}\;\left[\,
\binom{2n}{2k}\: -\:\frac{2k(n-k)}{n-1}\:\binom{n}{k}\,\right]\; x^k \, .
\end{eqnarray}
In all cases, the coefficients of the polynomials
are rather simple expressions in terms of binomial coefficients
\begin{equation}
\binom{n}{k} \; =\; \frac{n!}{k!\, (n-k)!} \: .
\end{equation}
The results for the graphs $A_n$ ($n\geq 1$) and $D_n$ ($n\geq 4$) are just 
those contained 
in \cite{ConSlo}, and the polynomials for $C_n$ ($n\ge 2$) can be derived
by the methods outlined there, if one observes that the {\em longer\/} 
roots of $C_n$ 
generate a sublattice that is equivalent to $\ZZ_{}^n$. 
However, the expressions for 
$B_n$ ($n\ge 2$), as those for $D_n$ ($n\ge 4$) here and in \cite{ConSlo},
are conjectures based on enumeration of coordination sequences for 
a large number of examples. The striking similarity between $B_n$ and $D_n$
might actually help to find a proof. The connection is rather intimate: while
all roots of $B_n$ generate $\ZZ_{}^n$, the long roots alone generate $D_n$,
and each point of $B_n$ can be reached from 0 by using a path with at most
one short root.

For the three root graphs related to the exceptional (simply laced) 
Lie algebras $E_6$, $E_7$, and $E_8$, the coordinator polynomials read
\begin{eqnarray}
P_{E_6}^{}(x)&=&1+66\, x+645\, x^2+1384\, x^3+645\, x^4+66\, x^5+x^6 \\
P_{E_7}^{}(x)&=&1+119\, x+2037\, x^2+8211\, x^3+8787\, x^4+
                2037\, x^5\nonumber\\*
             & &\;\;\mbox{} +119\, x^6+x^7\\
P_{E_8}^{}(x)&=&1+232\, x+7228\, x^2+55384\, x^3+133510\, x^4+
                107224\, x^5\nonumber\\*
             & &\;\;\mbox{} +24508\, x^6+232\, x^7+x^8
\end{eqnarray}
as has been proved in \cite{ConSlo}. Finally, for the two remaining root 
graphs we find
\begin{eqnarray}
P_{F_4}^{}(x) & = & 1+44\, x+198\, x^2+140\, x^3+x^4 \\
P_{G_2}^{}(x) & = & 1+10\, x+7\, x^2 \: .
\end{eqnarray}
Let us give an explicit proof for the last example. Clearly,
$s^{}_{G_2}(0)=1$ and $s^{}_{G_2}(1)=12$. Then, for $n\geq 2$,
one can explicitly show that the graph $G_2$, in comparison to $A_2$
(which also happens to be equivalent to the lattice generated by the 
long roots of $G_2$, see Figure~1),
has a shell structure with $s^{}_{G_2}(n) = 2 s^{}_{A_2}(n) + 
s^{}_{A_2}(n-1) = 18 n - 6$, from which the above statement follows.
By similar arguments, the other examples with short and long root 
vectors can be traced back to the lattice case; 
the corresponding shell structure 
of the root graph\clearpage is defined by the coordination 
spheres of its sublattice 
generated by the set of long root vectors.

{}Finally, it is interesting to note that the coordination sequences for
root graphs $\Lambda$ of type $A_n$, $C_n$, $D_n$, and $E_6$ 
result in self-reciprocal polynomials $P^{}_{\Lambda}(x)$, i.e.,
\begin{equation}
    P^{}_{\Lambda}(x) \; = \; x^n \cdot P^{}_{\Lambda}(1/x)
\end{equation}
while the others do not; for a geometric meaning of this property
we refer to \cite{ConSlo}.

\section*{Outlook}

We presented the coordination sequences and their generating functions
for root lattices and, more generally, graphs based upon the root systems,
namely for the series $A_n$ ($n\geq 1$), $B_n$ ($n\geq 2$), $C_n$ ($n\geq 2$), 
and $D_n$ ($n\geq 4$), and for the exceptional cases $E_6$, $E_7$, $E_8$, 
$F_4$, and $G_2$. Proofs of various cases can be found in 
Conway and Sloane \cite{ConSlo} 
or directly based on their results, but the generating functions for
$B_n$ and $D_n$ are still conjectural at the moment.

\section*{Acknowledgements}

We are grateful to N.~J.~A.~Sloane for sending us Ref.~\cite{ConSlo}
prior to publication.


\begin{thebibliography}{99}
\begin{small}

\bibitem{ConSlo}
Conway, J.~H.; Sloane, N.~J.~A.:
Low-dimensional lattices VII: coordination sequences.
To appear in: Proc.\ R.\ Soc.\ {\bf A} (1997).

\bibitem{GalMau1}
Galam, S.; Mauger, A.: 
Universal formulas for percolation thresholds.
Phys.\ Rev.\ {\bf E~53} (1996) 2177--2181.

\bibitem{GalMau2}
Galam, S.; Mauger, A.: 
A quasi-exact formula for Ising critical temperature 
on hypercubic lattices. 
Physica {\bf A~235} (1997) 573--576.

\bibitem{Bri}
Briggs, K.:
Self-avoiding walks on quasilattices.
Int.\ J.\ Mod.\ Phys.\ {\bf B~7} (1993) 1569--1575.

\bibitem{Laby}
Baake, M.; Grimm, U.; Baxter, R.~J.:
A critical Ising model on the Labyrinth.
Int.\ J.\ Mod.\ Phys.\ {\bf B~8} (1994) 3579--3600.

\bibitem{LY}
Simon, H.; Baake, M.; Grimm, U.:
Lee-Yang zeros for substitutional systems.
In: {\it Proceedings of the 5th International Conference on Quasicrystals\/}
(Eds.\ C.~Janot, R.~Mosseri), p.~100--103. 
World Scientific, Singapore 1995.

\bibitem{CS}
Conway, J.~H.; Sloane, N.~J.~A.:
Sphere Packings, Lattices and Groups (2nd ed.).
Springer, New York 1993.

\bibitem{Humphreys}
Humphreys, J.~E.:
Introduction to Lie Algebras and Representation Theory.
Springer, New York 1972. And:
Reflection Groups and Coxeter Groups. CUP, Cambridge 1990.

\bibitem{Keefe91}
O'Keeffe, M.:
$N$-Dimensional Diamond, Sodalite and Rare Sphere Packings.
Acta Cryst.\ {\bf A 47} (1991) 748--753.

\bibitem{Keefe95}
O'Keeffe, M.:
Coordination sequences for lattices.
Z.\ Kristallogr.\ {\bf 210} (1995) 905--908.

\bibitem{SloPlo95}
Sloane, N.~J.~A.; Plouffe, S.:
The Encyclopedia of Integer Sequences.
Academic Press, San Diego 1995.

\bibitem{Slo}
Sloane, N.~J.~A.:
An on-line version of the encyclopedia of integer sequences.
Electronic J.\ Combinatorics {\bf 1} (1994).

\bibitem{Wilf}
Wilf, H.~S.:
Generatingfunctionology (2nd ed.). Academic Press, Boston 1994.

\end{small}
\end{thebibliography}
\end{document}